\documentclass[conference]{IEEEtran}

\usepackage{amsmath, amssymb, bm, cite, epsfig, psfrag}
\usepackage{epstopdf}
\usepackage{graphicx}
\usepackage{algorithm,algpseudocode}
\usepackage{array}
\usepackage[margin=10pt,font=small]{caption}
\usepackage{bbm}
\usepackage{multirow}
\usepackage[usenames,dvipsnames]{xcolor}

\usepackage{subfigure}
\usepackage{etoolbox}
\usepackage{pbox}
\usepackage[width=0.48\textwidth]{caption}
\usepackage{colortbl}
\usepackage{fancyhdr}
\usepackage{tikz}
\usetikzlibrary{calc}

\newtoggle{conference}
\togglefalse{conference} % Use for arxiv version -- also change documentclass
\graphicspath{{figures/}}

\def\beq{\begin{equation}}
\def\eeq{\end{equation}}
\def\beqa{\begin{eqnarray}}
\def\eeqa{\end{eqnarray}}
\def\beqan{\begin{eqnarray*}}
\def\eeqan{\end{eqnarray*}}

\def\tm1{t\! - \! 1}
\def\tp1{t\! + \! 1}

\usepackage{lipsum}
\usepackage{fancyhdr}
\usepackage{datetime}

\begin{document}

\pagestyle{empty}

\title{3-D Statistical Channel Model for Millimeter-Wave Outdoor Mobile Broadband Communications}

\author{
	\IEEEauthorblockN{Mathew K. Samimi and Theodore S. Rappaport\\NYU WIRELESS, NYU Polytechnic School of Engineering\\mks@nyu.edu, tsr@nyu.edu}
}

\maketitle

%---------------------- Document Header --------------------------%
\begin{tikzpicture}[remember picture, overlay]
\node at ($(current page.north) + (0,-0.25in)$) {M. K. Samimi, T. S. Rappaport, ''3-D Statistical Channel Model for Millimeter-Wave Outdoor Mobile Broadband Communications,''};
\node at ($(current page.north) + (0,-0.4in)$) {\textit{accepted at the 2015 IEEE International Conference on Communications (ICC)}, 8-12 June, 2015.};
\end{tikzpicture}

\begin{abstract}
This paper presents an omnidirectional spatial and temporal 3-dimensional statistical channel model for 28 GHz dense urban non-line of sight environments. The channel model is developed from 28 GHz ultrawideband propagation measurements obtained with a 400 megachips per second broadband sliding correlator channel sounder and highly directional, steerable horn antennas in New York City. A 3GPP-like statistical channel model that is easy to implement in software or hardware is developed from measured power delay profiles and a synthesized method for providing absolute propagation delays recovered from 3-D ray-tracing, as well as measured angle of departure and angle of arrival power spectra. The extracted statistics are used to implement a MATLAB-based statistical simulator that generates 3-D millimeter-wave temporal and spatial channel coefficients that reproduce realistic impulse responses of measured urban channels. The methods and model presented here can be used for millimeter-wave system-wide simulations, and air interface design and capacity analyses.
\end{abstract}

\section{Introduction}

Millimeter-waves (mmWave) are a viable solution for alleviating the spectrum shortage below 6 GHz, thereby motivating many recent mmWave outdoor propagation measurements designed to understand the distance-dependent propagation path loss, and temporal and spatial channel characteristics of many different types of environments ~\cite{Rap13:2,MacCartney14:1,Roh14,Rap15,MacCartney15}. The mmWave spectrum contains a massive amount of raw bandwidth and will deliver multi-gigabit per second data rates for backhaul and fronthaul applications, and to mobile handsets in order to meet the expected 10,000x demand in broadband data over the next 10 years~\cite{Rap15}\cite{Pi11}. 

MmWave statistical spatial channel models (SSCMs) do not yet exist, but are required to estimate channel parameters such as temporal multipath delays, multipath powers, and multipath angle of arrival (AOA) and angle of departure (AOD) information. Both directional and omnidirectional channel models are needed based on real-world measurements. Further,  SSCMs are needed to carry out link-level and system-level simulations for analyzing system performance required for designing next-generation radio-systems. New channel modeling frameworks, modulation schemes and corresponding key requirements are currently being considered to address 5G network planning~\cite{Akdeniz14,Ghosh14,Samimi14}, and future directional on-chip antennas~\cite{Gutierrez09} require that 3-D statistical channel models be used at mmWave bands. This paper presents the world\rq{}s first 3-D SSCM based on wideband New York City measurements at 28 GHz.

Previous results obtained from the extensive New York City propagation database yielded directional and omnidirectional path loss models in dense urban line of sight (LOS) and non-line of sight (NLOS) environments~\cite{MacCartney14:2}, important temporal and spatial channel parameters, and distance-dependent path loss models at 28 GHz and 73 GHz based on measurements and ray-tracing results~\cite{Samimi14}\cite{Thomas14}. Initial MIMO network simulations were carried out in~\cite{Sun14} using a 2-dimensional (2-D) wideband mmWave statistical simulator developed from 28 GHz wideband propagation measurements~\cite{Samimi14}, and showed orders of magnitude increase in data rates as compared to current 3G and 4G LTE using spatial multiplexing and beamforming gains at the base station for both LOS and NLOS dense urban environments.

Statistical channel modeling methods have thus far focused on extracting models from measured power azimuth spectra and by modeling the elevation dimension using 3-dimensional (3-D) ray-tracing predictions in order to make up for the lack of measured elevation data~\cite{Thomas14} or to estimate 3-D angles in the absence of directional measurements due to the use of quasi-omnidirectional antennas~\cite{MIWEBA}. In this paper, we present a 3-D statistical spatial and temporal channel model for the  urban NLOS environment based on 28 GHz ultrawideband propagation measurements that used 3-D antenna positioning and directional antennas in New York City, where AOA elevation characteristics at the receiver have been extracted from measured power angular spectra (obtained with field measurements without the use of ray-tracing). This work extends our 2-D SSCM presented in~\cite{Samimi14} to a true 3-D model that includes the AOA of arriving multipath, as well as AOD, and mutipath time statistics.

\section{28 GHz Propagation Measurements}

In 2012, an ultrawideband propagation measurement campaign was carried out at 75 transmitter (TX) - receiver (RX) locations in New York City to investigate 28 GHz wideband channels using a 400 megachips per second (Mcps) broadband sliding correlator channel sounder and 24.5 dBi (10.9$^{\circ}$ half-power beamwidth (HPBW)) highly directional steerable horn antennas at the TX and RX, in dense urban LOS and NLOS environments~\cite{Rap13:2}. Over 4,000 power delay profiles (PDPs) were collected at unique azimuth and elevation pointing angles at both the TX and RX to properly describe mmWave wideband multipath channels over an 800 MHz RF null-to-null bandwidth in the time and spatial domains. Additional details pertaining to the 28 GHz measurement campaign and hardware equipment used can be found in~\cite{Rap13:2}~\cite{Samimi14}~\cite{MacCartney14:2}~\cite{Deng14}.

\section{Synthesizing 3-D Absolute Timing Omnidirectional Power Delay Profiles}

\begin{figure}[t]
    \begin{center}
        \includegraphics[width=3.5in]{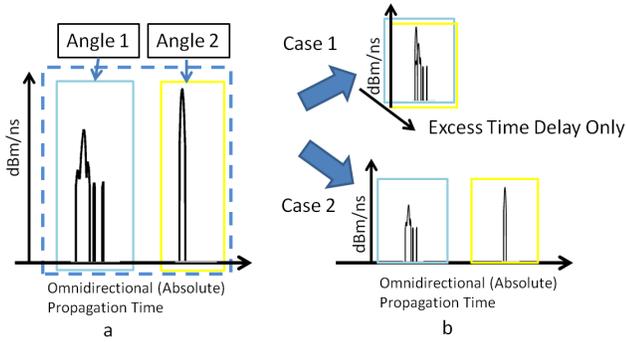}
    \end{center}
    \caption{Superimposed PDP of two individual received PDPs, where each PDP comes from a different AOA at the same RX location. The multipath signals from Angle 1 arrived before those of Angle 2 (i.e. multipath arriving at different times from two distinct lobes). The absolute propagation times were found using ray-tracing, thus allowing alignment with absolute timing of multipath signals originally measured at the RX, independent of AOAs (see~\cite{Samimi14}). }
    \vspace{-0.4cm}
    \label{fig:47}
    \end{figure}

The collected directional PDPs did not make use of absolute timing synchronization between the TX and RX over the various sweeps, and therefore provided received power over \textit{excess}, and not absolute, time delay. As illustrated in Fig.~\ref{fig:47}a, two typical measured excess time delay PDPs are shown as measured at two distinct AOA azimuth angles. The RX used the strongest arriving multipath component to trigger and establish the relative $t = 0$ ns time marker for all recorded PDPs, as illustrated in Case 1 of Fig.~\ref{fig:47}b where both PDPs from different angles are shown to start at $t = 0$ ns, and thus could not capture absolute arrival time of power. Case 2 illustrates the desired result, showing the two PDPs measured along an absolute propagation time axis, where \text{$t = 0$} ns corresponds to the energy leaving the TX antenna. In Case 2, the recorded excess time delay PDPs have been shifted in time appropriately, accounting for the propagation distance travelled of the first arriving multipath in each PDP, enabling accurate characterization of received power in time at the RX over all measured angles. 

\begin{figure}[t]
    \centering
 \includegraphics[width=3.5in]{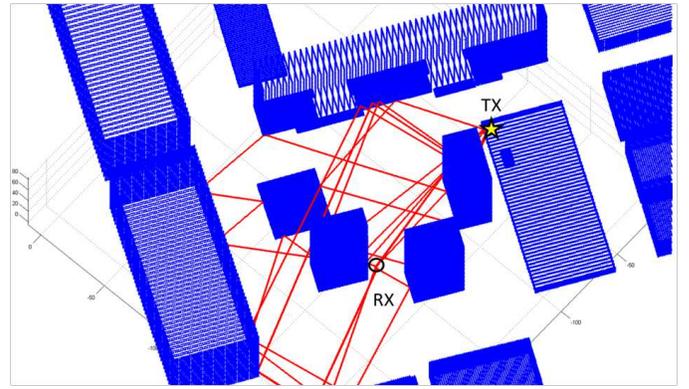}
    \caption{A 3-dimensional view of the downtown Manhattan area using the MATLAB-based 3-D ray-tracer. The rays which leave the TX and successfully arrive at the RX are shown in red, and represent multipath signal paths. The TX was located on the rooftop of the Coles Sports Center 7 m above ground (yellow star), and the RX was located 113 m away, 1.5 m above ground (black circle)~\cite{Rap13:2}.}
    \label{fig:RayTracedMap}
\end{figure}

We developed a MATLAB-based 3-D ray-tracer to predict the propagation time delays of the first arriving multipath components at the strongest measured AOAs for each TX-RX measured location. Ray-tracing techniques have previously been shown to provide accurate time and amplitude predictions~\cite{Rap15}\cite{Thomas14}\cite{Durgin97:1}. Fig.~\ref{fig:RayTracedMap} shows a typical ray-traced measured location where viable propagation paths between TX and RX are shown in red. The ray-tracing results showed strong spatial correlation with the measurements when comparing the strongest measured and predicted directions of departure and arrival (within $\pm$ two antenna beamwidths). The predicted propagation distances were paired with the closest strongest measured AOAs, and we superimposed each PDP recorded at the strongest AOAs in azimuth and elevation on an absolute propagation time axis by appropriately shifting and summing each PDP in time using the ray-tracing absolute time predictions. This method was performed over all measured RX locations, and allowed us to synthesize omnidirectional absolute timing PDPs as would have been measured with a quasi-omnidirectional isotropic antenna, yielding one unique omnidirectional 3-D power delay profile at each TX-RX location combination. Our measurement approach enabled high gain antennas to measure the channel over large distances in a directional mode, and the orthogonal beam patterns in adjacent beam pointing angles allow the orthogonal PDPs to be summed in time and space to form a 3-D omnidirectional model.

\section{3-D Statistical Channel Parameters}

Our statistical channel model recreates omnidirectional channel impulse responses $h(t,\overrightarrow{\mathbf{\Theta}},\overrightarrow{\mathbf{\Phi}})$, where $t$ denotes absolute propagation time delay, $\overrightarrow{\mathbf{\Theta}}=(\theta,\phi)_{TX}$ represents the vector of AOD azimuth and elevation angles, and $\overrightarrow{\mathbf{\Phi}}=(\theta,\phi)_{RX}$ is the vector of AOA azimuth and elevation angles, as defined in Eqs.~(\ref{IREq}), (\ref{3DEq1}), and (\ref{3DEq2}). Previous work modeled the channel impulse response as a function of time~\cite{Saleh87},  azimuth angle of arrival~\cite{Spencer97}\cite{Ertel98}, and AOD/AOA elevation and azimuth dimensions~\cite{WINNERII}. Note that~\cite{WINNERII} includes AOD and AOA elevation and azimuth information for in-building, indoor-to-outdoor, and outdoor-to-indoor scenarios, but not for the outdoor urban microcellular environment as considered in this paper. Here, we generalize the channel impulse response as a function of time, as well as a function of AOD and AOA azimuth/elevation angles, allowing realistic simulations of directional transmissions at both the TX and RX in azimuth and elevation dimensions. The channel impulse response between TX and RX is written as:
\begin{equation}\label{IREq}\begin{split}
h(t,\overrightarrow{\mathbf{\Theta}},\overrightarrow{\mathbf{\Phi}}) =& \sum_{l_1=1}^{L_{AOD}}\sum_{l_2=1}^{L_{AOA}} \sum_{n=1}^{N} \sum_{m=1}^{M_n}  a_{m,n,l_1,l_2} e^{j\varphi_{m,n}} \\ 
&\cdot \delta(t - \tau_{m,n}) \cdot \delta(\overrightarrow{\mathbf{\Theta}}-\overrightarrow{\mathbf{\Theta}}_{l_1}) \cdot \delta(\overrightarrow{\mathbf{\Phi}}-\overrightarrow{\mathbf{\Phi}}_{l_2})
\end{split}\end{equation}

\noindent where $L_{AOD}$ and $L_{AOA}$ are the number of AOD and AOA spatial lobes (defined in~\cite{Samimi14}\cite{Samimi13:1}, and in Section~\ref{sec:proc}), respectively; $N$ and $M_n$ are the number of time clusters and the number of intra-cluster subpaths in the $n$\textsuperscript{th} time cluster (as defined in Section~\ref{sec:cluster}), respectively; $a_{m,n,l_1,l_2}$ is the amplitude of the  $m$\textsuperscript{th} subpath component belonging to the $n$\textsuperscript{th} time cluster, departing from AOD lobe $l_1$, and arriving at AOA lobe $l_2$; $\varphi_{m,n}$, and $t_{m,n}$ are the phase and the propagation time of arrival of the $m$\textsuperscript{th} subpath component belonging to the $n$\textsuperscript{th} time cluster, respectively; $\overrightarrow{\mathbf{\Theta}}_{l_1}$ and $\overrightarrow{\mathbf{\Phi}}_{l_2}$ are the azimuth/elevation AODs and AOAs of lobes $l_1$ and $l_2$, respectively. In our channel model, each multipath component is assigned one joint AOD-AOA lobe combination, per the time cluster definition given in Section~\ref{sec:cluster}.  

Our statistical channel model also produces the joint AOD-AOA power spectra in the azimuth and elevation domains in 3-D, based on our 28 GHz New York City field measurements that only used a single TX pointing elevation angle at a $10^{\circ}$ downtilt and three RX elevation planes of $0^{\circ}$, $\pm 20^{\circ}$ (note: measurements at 73 GHz provided multiple AOD and AOA elevation angles, and may be extrapolated and used here). The spatial distribution of power is obtained by taking the magnitude squared of $h(t,\overrightarrow{\mathbf{\Theta}},\overrightarrow{\mathbf{\Phi}}) $, and integrating over the time dimension, as shown in (\ref{3DEq1}) and (\ref{3DEq2}):
\begin{align}\label{3DEq1}
P(\overrightarrow{\mathbf{\Theta}},\overrightarrow{\mathbf{\Phi}}) &= \int_{0}^{\infty} |h(t,\overrightarrow{\mathbf{\Theta}},\overrightarrow{\mathbf{\Phi}}) |^2 dt\\
\begin{split}P(\overrightarrow{\mathbf{\Theta}},\overrightarrow{\mathbf{\Phi}}) &= \sum_{l_1=1}^{L_{AOD}}\sum_{l_2=1}^{L_{AOA}}\sum_{n=1}^{N} \sum_{m=1}^{M_n} |a_{m,n,l_1,l_2}|^2  \\ \label{3DEq2}
&\cdot \delta(\overrightarrow{\mathbf{\Theta}}-\overrightarrow{\mathbf{\Theta}}_{l_1}) \cdot \delta(\overrightarrow{\mathbf{\Phi}}-\overrightarrow{\mathbf{\Phi}}_{l_2})
\end{split}
\end{align}

\noindent Note that $P(\overrightarrow{\mathbf{\Theta}},\overrightarrow{\mathbf{\Phi}})$ in Eq.~(\ref{3DEq2}) are the total received powers (obtained by integrating the PDP over time) assigned to the lobe AODs and AOAs.

\subsection{28 GHz Omnidirectional NLOS Path Loss Model}

The 28 GHz omnidirectional NLOS path loss model was recovered by summing the received powers measured at each and every azimuth and elevation unique pointing angle combination to recover the total received omnidirectional power at each TX-RX location~\cite{MacCartney14:2}. This procedure is valid since adjacent angular beamwidths are orthogonal to each other, and phases of arriving multipath components with different angular propagation paths can be assumed identically and independently distributed (i.i.d.) and uniform between 0 and $2\pi$~\cite{Rap15}, such that powers in adjacent beam angles can be added. After removing antenna gains and carefully removing double counts occuring from the TX and RX azimuth sweeps, we recovered the corresponding path loss at all measured NLOS locations, and extracted the path loss exponent and shadow factor using the 1 m close-in free space reference distance path loss model~\cite{Rap15}\cite{MacCartney14:2}:
\begin{equation}
PL_{NLOS}(d)[dB] = 61.4 + 34 \log_{10} (d) + \chi_{\sigma}, \hspace{.2cm}d > 1 \text{ m}
\end{equation}
\noindent where $\chi_{\sigma}$ is the lognormal random variable with 0 dB mean and shadow factor $\sigma = 9.7$ dB, and 61.4 dB is 28 GHz free space path loss at 1 m.

\subsection{Cluster and Lobe Statistics}
\label{sec:cluster}
The temporal and spatial components of our SSCM are modeled by a \textit{time cluster} and \textit{spatial lobe}, respectively, and faithfully reproduce omnidirectional PDPs and power azimuth spectra~\cite{Samimi14}. Time clusters model a group of multipath components travelling closely in time over all directions, and can represent one or more spatial or angular directions within the same time epoch. Spatial lobes represent a small contiguous span of angles at the RX (TX) where energy arrives (departs) over a small azimuth and elevation angular spread. In our SSCM, multiple time clusters can arrive in one spatial lobe, and a time cluster can arrive over many spatial lobes within a small span of propagation time. Time clusters and spatial lobes statistics can be easily extracted from the propagation measurements to build a 3GPP-like statistical channel model, including simple extensions to the current 3GPP and WINNER models that account for cluster subpath delays and power levels~\cite{3GPP:1}. Time cluster and spatial lobe characteristics were illustrated in~\cite{Samimi14} (see Figs. 3 and 4). The measured data suggests that multiple clusters in the time domain arrive up to several hundreds of nanoseconds in excess time delay for arbitrary pointing angles, observable due to our 24.5 dBi high gain antennas. We note that multipath components within a time cluster, i.e.,  intra-cluster subpath components, were successfully used to model the indoor office environment based on wideband measurements~\cite{Saleh87}~\cite{Spencer97}. 

Key parameters that serve as inputs to the 3-D mmWave SSCM are referred to as \textit{primary statistics}, and have been identified and defined in~\cite{Samimi14}, and include the number of time clusters and the cluster power levels. \textit{Secondary statistics} describe statistical outputs of the SSCM, and include the RMS delay spread and RMS angular spreads which reflect second-order statistics. Secondary statistics provide a means of testing the accuracy of a statistical channel model and simulator over a large ensemble of simulated outputs.

\subsection{Time Cluster Partitioning}

In this work, omnidirectional PDPs were partitioned in time based on a 25 ns minimum inter-cluster void interval, by assuming that multipath components fall within time clusters that are separated by at least 25 ns in time span. Walkways or streets between building facades are typically 8 m in width (roughly 25 ns in propagation delay). The 25 ns inter-cluster void interval allowed us to resolve measured multipath channel dynamics in a simple, yet powerful way, offering a scalable clustering algorithm that can be modified by changing the inter-cluster void interval for arbitrary time resolution. As the minimum inter-cluster void interval is increased, the number of time clusters in a PDP is expected to decrease, while the number of intra-cluster subpaths must consequently increase. In turn, the fewer time clusters must be allocated a larger portion of the total received power, while the greater number of intra-cluster subpaths will receive a lesser amount of the cluster power. While the minimum inter-cluster void interval heavily affects the outcome of the temporal channel parameter statistics, we note that the RMS delay spread is the only time parameter that remains unchanged for arbitrary inter-cluster void interval, making it a good but not the only proper indicator for comparing the simulated PDP outputs with the ensemble of measured PDPs. The 25 ns minimum inter-cluster void interval of our model is comparable to the best/maximum measured time resolution (20 ns) of a single multipath component in 3GPP and WINNER models.

\begin{figure}[t]
    \centering
 \includegraphics[width=3.5in]{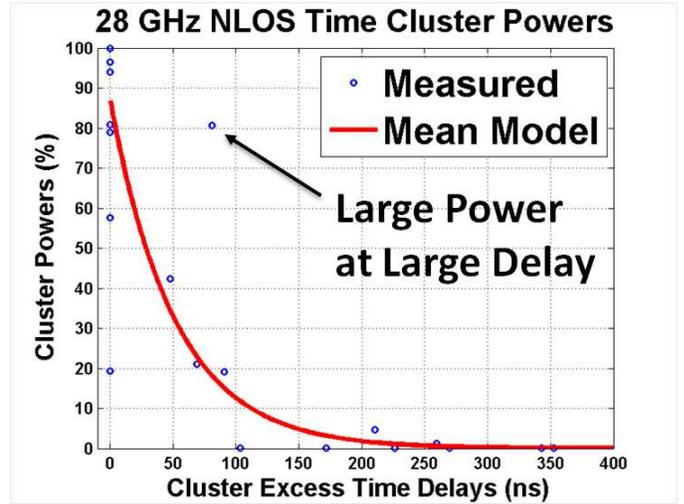}
    \caption{Temporal cluster powers normalized to omnidirectional total received power, over cluster excess delays, using a 25 ns minimum inter-cluster void interval. The superimposed least-squares regression exponential curve has an average cluster decay constant of $\Gamma=49.4$ ns, and a $y$-intercept of $\overline{P}_0=88.3$\% (see Step 7 in Section~\ref{sec:mmWaveProc}). A large cluster power can be seen at $\tau=80$ ns, which typically causes large delay spreads.}
    \label{fig:clusterPowers}
\end{figure}

\begin{figure}[t]
    \centering
 \includegraphics[width=3.5in]{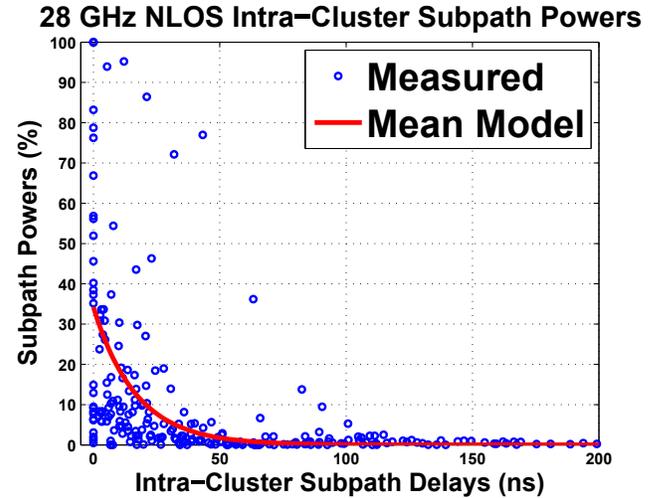}
    \caption{Intra-cluster subpath powers normalized to time cluster powers, using a 25 ns minimum inter-cluster void interval. The superimposed least-squares regression exponential curve has an average cluster decay constant of $\gamma=16.9$ ns, and a $y$-intercept of $\overline{P}_0=34.2$\% (see Step 8 in Section~\ref{sec:mmWaveProc}).}
    \label{fig:subpathPowers}
\end{figure}

Fig.~\ref{fig:clusterPowers} shows the temporal cluster powers normalized to the total received power in the omnidirectional profiles, and the least-squares regression exponential model that reproduces the measured cluster powers. The cluster powers were obtained by partitioning omnidirectional PDPs based on a 25 ns minimum inter-cluster void interval, finding the area under each time cluster, and dividing by the total power (area under the PDP). The mean exponential curve is parameterized using two parameters, the average cluster power $\overline{P}_0$ in the first received cluster (i.e., the $y$-intercept at $\tau=0$ ns), and the average cluster decay time $\Gamma$ defined as the time required to reach 37\% $(1/e)$ of $\overline{P}_0$. It is worth noticing the measured large cluster power at $\tau=80$ ns, containing close to 80\% of the total received power, corresponding to large fluctuations in cluster powers. This phenomenon causes large delay spreads, and is typically modeled using a lognormal random variable, as discussed in Step 5 of Section~\ref{sec:proc}. In Fig.~\ref{fig:clusterPowers}, we estimated $\overline{P}_0 = 0.883$, and $\Gamma=49.4$ ns. Similarly, Fig.~\ref{fig:subpathPowers} shows the intra-cluster subpath power levels (normalized to the total cluster powers), with $\overline{P}_0 = 0.342$ and $\gamma=16.9$ ns. The smaller subpath decay time physically means that intra-cluster subpaths decay faster than time clusters.

\subsection{3-D Lobe Thresholding}

The 3-D spatial distribution of received power is used to extract 3-D directional spatial statistics, by defining a \text{-10 dB} lobe threshold below the maximum peak power in the 3-D power angular spectrum, where all contiguous segment powers above this threshold in both azimuth and elevation were considered to belong to one 3-D spatial lobe. Spatial thresholding was performed on 3-D AOA power spectra, and 2-D AOD azimuth spectra, after linearly interpolating the directional measured powers (in linear units) to a 1$^{\circ}$ angular resolution in azimuth and elevation domains to enhance the angular resolution of our spatial statistics. 

\section{Generating 3-D mmWave Channel Coefficients}
\label{sec:proc}
A statistical channel model is now presented for generating 3-D mmWave PDPs and spatial power spectra that accurately reflect the statistics of the measurements over a large ensemble, valid for 28 GHz NLOS propagation with a noise floor of -100 dBm over an 800 MHz RF null-to-null bandwidth, and for a maximum system dynamic range of 178 dB~\cite{Rap13:2}. The clustering and lobe thresholding methodologies used in our work effectively de-couple temporal from spatial statistics. Step 12 of our channel model bridges the temporal and spatial components of the SSCM by randomly assigning temporal subpath powers to spatial lobe AODs and AOAs, thereby re-coupling the time and space dimensions to provide an accurate joint spatio-temporal SSCM. In the following steps, $DU$ and $DLN$ refer to the discrete uniform and discrete lognormal distributions, respectively, and the notation $[x]$ denotes the closest integer to $x$. Also, Steps 11 through 15 apply to both AOD and AOA spatial lobes.

\subsection{Step Procedure for Generating Channel Coefficients}
\label{sec:mmWaveProc}
\noindent \textit{Step 1: Generate the T-R separation distance $d$ ranging from 60-200 m in NLOS (based on our field measurements, and could be modified with further measurements)}:
\begin{equation}
d \sim Uniform(d_{min} = 60,d_{max}=200)
\end{equation}
\noindent Note: To validate our simulation, we used the above distance ranges, but for standards work any distance less than 200 m is valid.

\noindent \textit{Step 2: Generate the total received omnidirectional power $P_r$ (dBm) at the RX location using the 1 m close-in free space reference distance path loss model~\cite{Rap15}\cite{MacCartney14:2}}:
\begin{align}
&P_r [dBm] = P_t + G_t + G_r - PL[dB]\\
&PL_{NLOS}[dB] = 61.4 + 34\log_{10}\left(d\right)+\chi_{\sigma}, \hspace{.3cm}d\geq 1\text{ m}
\end{align}
\noindent where $P_t$ is the transmit power in dBm, $G_t$ and $G_r$ are the TX and RX antenna gains in dBi, respectively, $\lambda = 0.0107$ m, and $\overline{n}$ = 3.4 is the path loss exponent for omnidirectional TX and RX antennas~\cite{MacCartney14:2}. $\chi_{\sigma}$ is the lognormal random variable with 0 dB mean and shadow factor $\sigma=9.7$ dB.

\noindent \textit{Step 3: Generate the number of time clusters $N$ and the number of spatial AOD and AOA lobes $(L_{AOD}, L_{AOA})$ at the TX and RX locations, respectively:}
\begin{align}
&N \sim DU[1,6]\\
& L_{AOD} \sim \min \bigg\{ L_{max},\max \Big\{ 1, \min  \big\{A,N   \big\} \Big\} \bigg\}\\
& L_{AOA} \sim \min \bigg\{ L_{max}, \max \Big\{ 1, \min  \big\{B,N   \big\} \Big\} \bigg\}\\
\text{and: }& A\sim \text{Poisson}(\mu_{AOD}+0.2)\\
& B \sim \text{Poisson}(\mu_{AOA}+0.1)
\end{align}

\noindent where $\mu_{AOD}=1.6$ and $\mu_{AOA} = 1.7$ are the mean number of AOD and AOA lobes observed in Manhattan, respectively, and $L_{max}=5$ is the maximum allowable number of lobes, for both AODs and AOAs. Work in~\cite{Samimi13:1} found the mean number of AOA lobes to be 2.5 using a -20 dB threshold, whereas here we use a -10 dB threshold. The 28 GHz NLOS measurements found the maximum number of clusters $N_{max}=5$ based on measurements in~\cite{Samimi13:1}, while at 73 GHz we found $N_{max}=6$, therefore we used $N_{max}=6$ for both frequency bands. Note that $(L_{AOD},L_{AOA})$ must always remain less than or equal to $N$, since the number of spatial lobes must be at most equal to the number of traveling time clusters in the channel. Also, the use of coin flipping was introduced in our previous work~\cite{Samimi14} to generate the pair $(N,L_{AOA})$, to obtain a close fit between measured and statistical data. In this work, however, we use standard well-known distributions, without the use of coin flipping, to promote ease of use in simulated software.

\noindent \textit{Step 4: Generate the number of cluster subpaths (SP) $M_n$ in each time cluster:} 
\begin{align}
&M_n \sim DU[1, 30]\hspace{.5cm},\hspace{.5cm} n = 1, 2, ... N
\end{align}
\noindent At 28 GHz, the maximum and second to maximum number of cluster subpaths were 53 and 30, respectively, over all locations, while it was 30 at 73 GHz, so we choose to use 30 for both frequency bands.

\noindent \textit{Step 5: Generate the intra-cluster subpath excess delays $\rho_{m,n}$}: 
\begin{equation}
\rho_{m,n}(B_{bb}) =  \bigg\{ \frac{1}{B_{bb}}\times (m-1) \bigg\}^{1+X} 
\end{equation}
\noindent where  $B_{bb} = 400$ MHz is the baseband bandwidth of our transmitted PN sequence (and can be modified for different baseband bandwidths), $X$ is uniformly distributed between 0 and 0.43, and $m = 1,2,...M_n, n = 1,2,...N$. This step allows for a minimum subpath time interval of 2.5 ns, while reflecting our observations that the time intervals between intra-cluster subpaths tend to increase with time delay. The bounds on the uniform distribution for $X$ will likely differ depending on the site-specific environment, and can be easily adjusted to fit field measurement observations.

\noindent \textit{Step 6: Generate the cluster excess delays $\tau_n$ (ns):} 
\begin{align}
&\tau^{\prime\prime}_n \sim \text{Exp}(\mu_{\tau})\\
&\Delta \tau_n = \text{sort}(\tau^{\prime\prime}_n)-\min(\tau^{\prime\prime}_n) \label{sort}
\end{align}
\vspace{-.6cm}
\begin{align}
&\tau_n =
\begin{cases}  
      0, &n = 1\\		
    \tau_{n-1}+\rho_{M_{n-1},n-1}+\Delta \tau_n+ 25, &n = 2, ..., N 
\end{cases}
\end{align}

\noindent where $\mu_{\tau}=83$ ns, and \textit{sort()} in (\ref{sort}) orders the delay elements $\tau^{\prime\prime}_n$ from smallest to largest. This step assures no temporal cluster overlap by using a 25 ns minimum inter-cluster void interval.

\noindent \textit{Step 7: Generate the time cluster powers $P_n$ (mW):}
\begin{align}\label{scale}
&P^{\prime}_n = \overline{P}_0 e^{-\frac{\tau_n}{\Gamma}} 10^{\frac{Z_n}{10}}\\ \label{eq2}
&P_n = \frac{P^{\prime}_n}{\sum^{k=N}_{k=1} P^{\prime}_k}\times P_r [mW]\\ \label{eq3}
&Z_n \sim N(0,3\text{ dB} ),n = 1, 2, ... N
\end{align}
\noindent where $\Gamma=49.4$ ns is the cluster decay time, $\overline{P}_0=0.883$ is the average (normalized) cluster power in the first arriving time cluster, and $Z_n$ is a lognormal random variable with 0 dB mean and $\sigma=3$ dB. Eq.~(\ref{eq2}) ensures that the sum of cluster powers adds up to the omnidirectional power $P_r$, where $\overline{P}_0$ cancels out and can be used as a secondary statistic to validate the channel model. 

\noindent \textit{Step 8: Generate the cluster subpath powers $\Pi_{m,n}$ (mW) }: 
\begin{align}
&\Pi^{\prime}_{m,n} = \overline{\Pi}_0 e^{-\frac{\rho_{m,n}}{\gamma}}  10^{\frac{U_{m,n}}{10}}\\ \label{eqSP}
&\Pi_{m,n} = \frac{\Pi^{\prime}_{m,n}}{\sum^{k=N}_{k=1} \Pi^{\prime}_{k,n}}\times P_n [mW]\\
& U_{m,n} \sim N(0, 6\text{ dB})\\
&m = 1, 2, ..., M_n\hspace{.3cm},\hspace{.3cm} n = 1, 2, ... N
\end{align}
\noindent  where $\gamma=16.9$ ns is the subpath decay time, $\overline{\Pi}_0 = 0.342$ is the average subpath power (normalized to cluster powers) in the first intra-cluster subpath, and $U_{m,n}$ is a lognormal random variable with 0 dB mean and $\sigma= 6$ dB. For model validation, the minimum subpath power was set to -100 dBm. Eq. (\ref{eqSP}) ensures that the sum of subpath powers adds up to the cluster power. Note: our measurements have much greater temporal and spatial resolution than previous models. Intra-cluster power levels were observed to fall off exponentially over intra-cluster time delay, as shown in Fig.~\ref{fig:subpathPowers} and in~\cite{Samimi14}.

\noindent \textit{Step 9: Generate the cluster subpath phases $\varphi_{m,n}$ (rad) }:
\begin{align}
&\varphi_{m,n} = \varphi_{1,n} + 2\pi f \rho_{m,n}\\
& \varphi_{1,n} \sim U(0,2\pi)\\
& m = 2, ..., M_n, n = 1, 2, ..., N
\end{align}

\noindent where $f = 28 \times 10^{9}$ Hz, and $\rho_{m,n}$ are the intra-cluster subpath delays in $s$ from Step 5, where $f$ can be any carrier frequency. The subpath phases $\varphi_{m,n}$ are i.i.d and uniformly distributed between 0 and $2\pi$, as modeled in~\cite{Saleh87}~\cite{Spencer97}. 

\noindent \textit{Step 10: Recover absolute time delays $t_{m,n}$ (ns) of cluster subpaths using the T-R Separation distance:} 
\begin{align}
&t_{m,n} = t_0 + \tau_n + \rho_{m,n}\hspace{.5cm}, \hspace{.5cm}t_0 = \frac{d}{c}
\end{align}
\noindent where $m = 1,2,...M_n, n = 1,2,...N$, and $c = 3\times 10^8 \text{ } m/s$ is the speed of light in free space.

\noindent \textit{Step 11 a: Generate the mean AOA and AOD azimuth angles $\theta_i(^{\circ})$ of the 3-D spatial lobes to avoid overlap of lobe angles:}
\begin{align}
&\theta_{i} \sim DU[\theta_{min},\theta_{max}]\hspace{.3cm},\hspace{.3cm} i = 1, 2, ..., L\\
& \theta_{min} = \frac{360(i-1)}{L}\hspace{.4cm},\hspace{.3cm} \theta_{max}  = \frac{360i}{L}
\end{align}

\noindent \textit{Step 11 b: Generate the mean AOA and AOD elevation angles $\phi_i(^{\circ})$ of the 3-D spatial lobes:}
\begin{align}
&\phi_{i} \sim [N(\mu,\sigma)]\hspace{.3cm},\hspace{.3cm} i = 1, 2,..., L. 
\end{align}
\noindent Positive and negative values of $\phi_i$ indicate a direction above and below horizon, respectively. While our 28 GHz Manhattan measurements used a fixed 10$^{\circ}$ downtilt at the transmitter, and considered fixed AOA elevations planes of 0$^{\circ}$ and $\pm 20^{\circ}$ at the receiver, mmWave transmissions are expected to beamform in the strongest AOD and AOA elevation and azimuth directions as was emulated in our 73 GHz Manhattan measurements in Summer 2013~\cite{MacCartney14:1}. Thus, we specify $(\mu, \sigma)=(-4.9^{\circ},4.5^{\circ}$) for AOD elevation angles, and $(\mu,\sigma)=(3.6^{\circ},4.8^{\circ})$ for AOA elevation angles from our 73 GHz NLOS measurements.

\noindent \textit{Step 12: Generate the AOD and AOA lobe powers $P(\theta_i, \phi_i)$ by assigning subpath powers $\Pi_{m,n}$ successively to the different AOD and AOA lobe angles ($\theta_i,\phi_i$):}
\begin{align}
P(\theta_i, \phi_i) =& 	\sum_{n=1}^N\sum_{m=1}^{M_n} \delta_{i w_{m,n}} \Pi_{m,n}\hspace{.3cm},\hspace{.3cm} i = 1, 2, ..., L\\
&w_{m,n} \sim DU[1,L] \text{ and }\delta_{rs} = 
\begin{cases}
0, r \neq s\\
1, r = s
\end{cases}
\end{align}

\noindent where $\delta_{rs}$ corresponds to the Kronecker delta. The cluster subpath $(m,n)$, with power level $\Pi_{m,n}$, is assigned to lobe $i$ only if $w_{m,n}=i$. This step distributes subpath powers into the spatial domains based on measurements in~\cite{Samimi13:1}.

\noindent \textit{Step 13: Generate the AOA and AOD lobe azimuth and elevation spreads $K_i$ (azimuth) and $H_i$ (elevation):}

\noindent For AODs: \text{$K_i\sim \max(5^{\circ},[N(30^{\circ}, 16^{\circ})])$, $H_i = 10^{\circ}$}\\
\noindent For AOAs: \text{$K_i\sim DLN(32^{\circ}, 18^{\circ})$, $H_i\sim \max(5^{\circ}, [N(31^{\circ}, 11^{\circ})])$}

\noindent AOD elevation spreads are fixed at $10^{\circ}$ based on our 28 GHz measurements that used a 10.9$^{\circ}$ antenna beamwidth. We have allowed for at most $10\%$ lobe azimuth and elevation overlap in adjacent spatial lobes.

\noindent \textit{Step 14: Generate the discretized lobe segment azimuth and elevation angles ($\theta_{i,j}, \phi_{i,l}$) for lobe AODs and AOAs:}
\begin{align}
\theta_{i,j} = \theta_i+k_j, j = 1,2,...K_i, i=1,2,...L\\
\phi_{i,l} = \phi_i+h_l, l = 1,2,...H_i, i=1,2,...L
\end{align}

\noindent where: $(X, W) \sim DU[0,1]$, $Y=1-X$, $Z=1-W$.
\begin{equation}
   \begin{cases}
k_j=-\frac{K_i-1}{2},...,-1,0,1,...,\frac{K_i-1}{2},& K_i \text{ odd}\\
k_j=-\frac{K_i}{2}+Y,...,-1,0,1,..,\frac{K_i}{2}-X,& K_i \text{ even}\\
\end{cases}
\end{equation}
\begin{equation}
   \begin{cases}
h_l=-\frac{H_i-1}{2},...,-1,0,1,...,\frac{H_i-1}{2},& H_i \text{ odd}\\
h_l=-\frac{H_i}{2}+W,...,-1,0,1,...,\frac{H_i}{2}-Z,& H_i \text{ even}\\
\end{cases}
\end{equation}

\noindent This step discretizes the spatial lobes into 1$^{\circ}$ angular segments in both azimuth and elevation dimensions.

\noindent \textit{Step 15: Generate the AOD and AOA lobe angular powers $P(\theta_{i,j},\phi_{i,l})(mW)$ at each $1^{\circ}$ angular segment:} 
\begin{align}
&P(\theta_{i,j},\phi_{i,l}) = R(\Delta \theta_{i,j},\Delta \phi_{i,l}) P(\theta_i,\phi_i)\\
&R(\Delta \theta_{i,j},\Delta \phi_{i,l}) = \max \bigg\{ e^{-\frac{1}{2}\left(\frac{(\Delta \theta_{i,j})^2}{\sigma_{\theta_{i}}^2}+\frac{(\Delta \phi_{i,l})^2}{\sigma_{\phi_{i}}^2}\right) }, \frac{1}{10} \bigg\}
\end{align}
\vspace{-.65cm}
\begin{align}
& j = 1, 2, ...,K_i, l = 1, 2, ..., H_i, i = 1,2,...,L
\end{align}

For AODs, $\sigma_{\theta_i}\sim N(6.6^{\circ},3.5^{\circ})$ and $\sigma_{\phi_i}\sim N(5^{\circ},0^{\circ})$.

For AOAs, $\sigma_{\theta_i}\sim N(6^{\circ},1^{\circ})$ and $\sigma_{\phi_i}\sim N(6^{\circ},2^{\circ})$.

\subsection{Measured vs. Simulated Statistics using a MATLAB-Based Statistical Simulator}

\begin{figure}[t]
    \centering
 \includegraphics[width=3.5in]{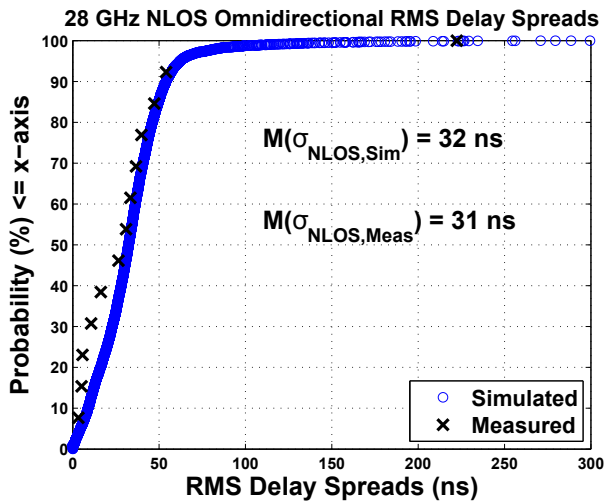}
    \caption{28 GHz NLOS synthesized and simulated omnidirectional RMS delay spreads. The median RMS delay spreads were 31 ns and 32 ns, for the synthesized and simulated data sets, respectively. The close agreement between measured and simulated data validates our NLOS statistical channel model.}
    \label{fig:CDFOmni}
\end{figure}

\begin{figure}[t]
    \centering
 \includegraphics[width=3.5in]{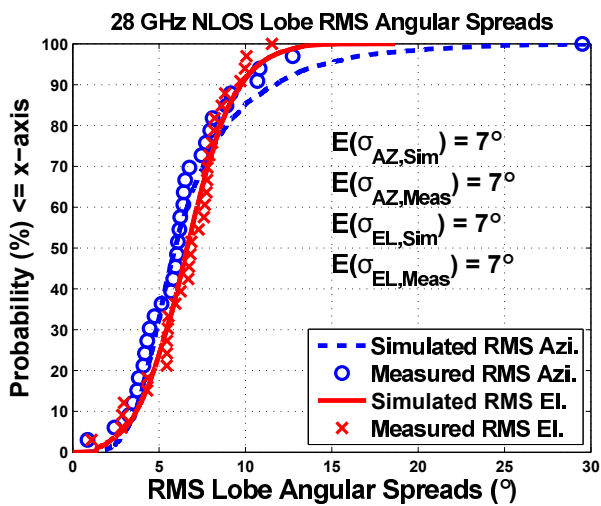}
    \caption{28 GHz NLOS RMS lobe azimuth and elevation spreads, measured as compared to simulated, using a -10 dB lobe threshold. The simulated data is in good agreement to the measured RMS angular spreads, validating the spatial component of our NLOS statistical channel model.}
    \label{fig:CDFAngular}
\end{figure}

A MATLAB-based statistical simulator using our 3-D SSCM generated a large ensemble (10,000) of mmWave PDPs, and AOD and AOA power spectra, where simulated and measured channel statistics were compared. Fig.~\ref{fig:CDFOmni} shows the cumulative distribution function (CDF) of the synthesized and simulated RMS delay spreads, showing relatively close agreement, with a median of 31 ns and 32 ns, respectively. The RMS delay spread CDF was skewed due to one large value of 222.4 ns, and so the median was chosen to reflect the empirical distribution trend. Fig.~\ref{fig:CDFAngular} shows the CDF of the RMS lobe azimuth and elevation spreads from the measured and simulated AOA power spectra, showing very good agreement to field measurements with identical measured and simulated means of 7$^{\circ}$ over both azimuth and elevation. 

\section{Conclusion}

This paper presents the first comprehensive 3-D statistical spatial channel model for mmWave NLOS communication channels. The thousands of measured PDPs have allowed us to create a 3-D SSCM that recreates the measured channel statistics over a large ensemble of simulated channels, and can be extended to arbitrary bandwidths and antenna patterns for use in physical layer simulations, such as physical layer design, and 3-D beamforming and beamcombining simulations used in MIMO systems in NLOS environments.
\bibliographystyle{IEEEtran}
\bibliography{ICC15_MKS}
\end{document}